\documentstyle[11pt]{article}
\input epsf
\begin{document}
\addtolength{\baselineskip}{0.25\baselineskip}
\thispagestyle{empty}
\rightline{CU-TP-668}
\vskip 2cm
\centerline{\Large\bf Self-Dual Anyons in Uniform Background Fields}

\vskip 1.5cm
\centerline{{\large\it Kimyeong Lee} \footnote{e-mail address:
klee@cuphyh.phys.columbia.edu} {\it and Piljin Yi} \footnote{e-mail address:
piljin@cuphyb.phys.columbia.edu }}
\vskip 3mm
\centerline{ Physics Department, Columbia University}
\centerline{ New York, N.Y.  10027, U.S.A.}
\vskip 3cm
\centerline{\bf ABSTRACT}
\vskip 5mm
\begin{quote}{\small
We study relativistic self-dual Chern-Simons-Higgs systems in the presence
of uniform background fields that explicitly break CTP. A rich, but discrete
vacuum structure is found when the gauge symmetry is spontaneously broken,
while the symmetric phase can have an infinite vacuum degeneracy at tree
level. The latter is due to the proliferation of neutral solitonic states
that cost zero energy. Various novel self-dual solitons, such as these, are
found in both the symmetric and the asymmetric phases. Also by considering a
similar system on a two-sphere and the subsequent large sphere limit, we
isolate sensible and finite expressions for the conserved angular and linear
momenta, which satisfy anomalous commutation relations. We conclude with a
few remarks on unresolved issues.}
\end{quote}

\newpage

\section{Introduction}

In Abelian Maxwell-Higgs systems, there are well-known topologically
stable vortex solutions in three dimensions. With a specific coupling
constants, the energy functional can be saturated by the so-called
self-dual configurations satisfying certain first-order differential
equations \cite{Bogo}. When the kinetic term for the gauge field includes
Chern-Simons term, the corresponding self-dual models are also found
and studied \cite{Hong,Min,Clee}.

An interesting generalization of such self-dual models arises when the
system is coupled to an external background charge density or an external
magnetic field. In particular, Maxwell-Higgs systems with the uniform
external electric charge  density was argued to describe the real
superconductor more closely than ones without. The self-dual limit of
these systems has been investigated extensively by one of the authors
\cite{Klee}.  In this paper, we want to consider two more self-dual models,
which incorporate a Chern-Simons term. A nonrelativistic version with an
external magnetic field was studied before as an effective field theory
for the fractional quantum Hall effects \cite{Zhang}.

There are several novel features in (Maxwell) Chern-Simons-Higgs
systems with the background charge density. Their structure is much
richer than the systems without. First, homogeneous ground states are
possible in two rather different manners: either a symmetric phase
with a uniform magnetic field or an asymmetric phase with a uniform
Higgs charge density. (But they obviously belong to two different
superselection sectors, even if the spatial volume is finite.) Second,
a magnetoroton mode is possible in the asymmetric phase for certain
parameter ranges.  This mode has the lowest energy at a nonzero wavelength.
Third, such a roton mode, when its energy is imaginary, leads to the
instability of the homogeneous asymmetric vacuum. When this happens,
the resulting ground state may have a crystal structure in charge
density. (If the symmetric phase is unstable on the other hand, the
resulting ground state would be a vortex lattice.) Then the
translation symmetry must be spontaneously broken and there will be a
sound wave as Goldstone boson. Fourth, the CTP symmetry is explicitly
broken and so solitons and antisolitons have in general
different mass spectrums. The Lorentz symmetry is also broken and some
solitons can have zero or negative rest mass even though it must
have positive kinetic mass. Fifth, the angular momentum operator
from the Noether theorem should be modified.  Some of these features
are studied previously for different models \cite{Zhang,Donatis,Ezawa}.

In this paper, we concentrate on the {\it self-dual} Chern-Simons
Higgs models with a uniform background charge density. As we shall see
shortly, the usual extended supersymmetry associated with many
self-dual models is no longer manifest in the presence of the
background charge density.

We start by introducing self-dual models in Sec.[2], and then go on to
study the vacuum structure of the model in Sec.[3]. We study the
homogeneous symmetric vacuum that exists for all self-dual couplings,
and find it stable even when the mass of elementary excitations
are {\it imaginary}. The vacuum structure of the
asymmetric phase depends on the parameters of the model and turns out
to be quite rich. We repeat the stability analysis for each homogeneous
asymmetric vacuum.  Rather interesting species of self-dual solitons
appear in some of these vacua, which will be investigated in Sec.[4].
Here, we will find that the symmetric phase is in fact infinitely
degenerate for certain parameter range.
The section [5] is devoted to the matter of conserved angular
momentum. The conserved Noether angular momentum turns out to be
inappropriate due to severe divergences, and needs to be modified.  A
satisfactory and physically well-motivated finite expression is found
by considering the same system on a sphere. Some concluding remarks
are discussed in Sec.[6].

\section{Model}

The first self-dual model, which is the one we will study in detail
here, is the theory of a Higgs field $\phi= f e^{i\theta}/\sqrt{2}$
coupled to a Chern-Simons gauge field $A_\mu$. The gauge field is
coupled to a uniform background electric charge density $\rho_e$, so
that the Lagrangian of the model can be written as,
\begin{equation}{\cal L} =  {\kappa \over 2}
\epsilon^{\mu\nu\rho}A_\mu \partial_\nu A_\rho + |D_\mu \phi|^2
- U - \rho_e A_0 ,
\end{equation}
where the self-dual potential is given by a specific form:
\begin{equation}
U = \frac{1}{ 4\kappa^2}|\phi|^2(2|\phi|^2 - v^2)^2 -
\frac{\rho_e}{2\kappa} (2|\phi|^2 - v^2),
\end{equation}
with $D_\mu \phi = (\partial_\mu + iA_\mu)\phi$.  The parameters
$v^2$ and $\rho_e$ can be either positive or negative, but without loss of
generality we may assume that $\kappa>0$.  By shifting $A_i
\rightarrow A_i - \rho_e \epsilon_{ij} x^j /2\kappa$,
we can see that the background electric charge $\rho_e$ is
equivalent to a uniform background magnetic field $F^e_{12} =
\rho_e/\kappa$.  The C, P, and T are all broken: the  Chern-Simons term
breaks the P and T, while the background charge term breaks
the C and CTP transformation.

The second model is a Maxwell Chern-Simons Higgs theory with a
neutral scalar field $N$,  whose Lagrangian is
\begin{equation}
 \tilde{\cal L} = - \frac{1}{4e^2} F_{\mu\nu} F^{\mu\nu} +
 \frac{\kappa}{ 2}\epsilon^{\mu\nu\rho}A_\mu \partial_\nu A_\rho +
 \frac{1}{ 2e^2}(\partial_\mu N)^2 + |D_\mu \phi|^2 - \tilde{U} -
\rho_e A_0 ,
\end{equation}
where the potential is
\begin{equation}
\tilde{U} = N^2 |\phi|^2 + \frac{e^2}{8}( 2|\phi|^2-v^2 - 2\kappa N)^2 -
\rho_e N.
\end{equation}
Actually, the first model can be obtained from the second by taking
the limit $e^2\rightarrow \infty$. The Maxwell term and the kinetic
term for $N$ become negligible, while the field equation for $N$ may
be used to recover $\cal L$ from $\tilde{\cal L}$. Similar models
without the background charge density have been studied before \cite{Clee},
and also the pure Maxwell-Higgs system with $\kappa=0$ was studied
extensively by one of the authors \cite{Klee}.  Except for the case
$\kappa=0$ where the parity is a good symmetry, there is no qualitative
difference between the two models and so we will focus on the simpler
pure Chern-Simons-Higgs model $\cal L$.

\vskip 0.5cm
The system is invariant under the local gauge transformations.
Gauss's constraint obtained by varying $A_0$ is
\begin{equation}
 \kappa  F_{12} +  f^2 (\dot{\theta} + A_0 ) - \rho_{\rm e} = 0 ,
\label{Gauss}
\end{equation}
where the dot denotes the time derivative. The electric charge current
$-f^2(\partial_\mu \theta + A_\mu) $ is conserved and the corresponding
conserved charge is $Q= -\int d^2x f^2(\dot{\theta} +A_0) $.

Note that both the total electric charge and the total magnetic flux
must be conserved independently, satisfying Gauss's constraint. (One
can imagine the theory is on a spatially compact manifold so that
``total'' magnetic flux and ``total'' electric charge make sense.)
Therefore, we find infinitely many superselection sectors, each of
which is labeled by the total electric charge or magnetic flux of the
system. Later on, we shall consider cases where the background charge
density $\rho_e$ is cancelled at spatial infinity by
either magnetic flux $F_{12}$ or the Higgs charge density
$-f^2(\dot{\theta} +A_0) $.  Collectively, the first class of sectors
are to be called {\it the symmetric phase} and the latter {\it the
asymmetric phase}, respectively, for the obvious reason.

Since the background charge density is uniform, the space-time
translation symmetry is preserved. The corresponding energy-momentum
tensor $T_{\mu\nu}$ obtained through the Noether theorem is
\begin{equation}
T_{\mu\nu} = \frac{\kappa}{2} \epsilon_{\mu\rho\sigma} A^\sigma
\partial_\nu A^\rho + \partial_\mu f \partial_\nu f +
f^2(\partial_\mu \theta + A_\mu )\partial_\nu \theta - \eta_{\mu\nu}
{\cal L}.
\end{equation}
The conserved energy of the system is then $ E=\int d^2 x \, T_{00}. $
After a partial integration,  Gauss's constraint (\ref{Gauss})
can be used to show that
\begin{equation}
E= \int d^2x \biggl\{\frac{1}{2}\dot{f}^2 + \frac{1}{2} (\partial_i
f)^2 + \frac{1}{2} f^2(\dot{\theta} +A_0)^2 + \frac{1}{2} f^2(
\partial_i \theta +A_i)^2 + U \biggr\}.
\end{equation}
By using  Gauss's constraint (\ref{Gauss}) again and integrating
by parts, we rewrite the energy as
\begin{eqnarray}
E&=&\  \int d^2x \biggl\{\frac{1}{2} \dot{f}^2 + \frac{1}{2}
(\partial_i f + \epsilon_{ij} f (\partial_j \theta + A_j) )^2
+ \frac{1}{2} f^2( \dot{\theta} + A_0 - \frac{1}{2\kappa} (f^2-v^2))^2
\biggr\} \nonumber \\
&+& \frac{v^2}{2} \int d^2x F_{12} - \int d^2x\; \partial_i [\frac{1}{2}
\epsilon_{ij} f^2(\partial_j \theta + A_j)].  \label{energy}
\end{eqnarray}
The last term vanishes identically, as long as $F_{12}=0$ or $f^2=0$
asymptotically. Recall that these conditions are characteristic
of the asymmetric phase and the symmetric phase alluded to earlier.

In the asymmetric phase, we naturally consider the excited states of
finite total magnetic flux $\Psi = \int d^2x \,F_{12}$, and for those
configurations we find  the following self-dual bound,
\begin{equation}
E \ge \frac{v^2}{2} \Psi . \label{SDE}
\end{equation}
As we will see later, within certain parameter range, this bound can
be saturated even when $v^2\Psi $ is actually negative.

In the symmetric phase, however, there exists a finite average magnetic
flux density, $F^e_{12}=\rho_e /\kappa$. Then, the symmetric phase
comes with a nonzero vacuum energy, $E_0 =( v^2/2)\int F^2_{12}$, as
is evident in Eq. (\ref{energy}).
On top of this, of course, there will be localized net
fluxes of solitonic configurations.
We must then consider the excited states of finite net excess magnetic flux
$\Psi - \Psi_0 = \int d^2x \,( F_{12} - F^e_{12})$, the excess
energy of which is again bounded below
\begin{equation}
E -  E_0  \ge \frac{v^2}{2} (\Psi -\Psi_0) =- \frac{v^2}{2} \int d^2x
f^2(\dot{\theta} + A_0).
\end{equation}

In order to saturate these self-dual bounds, the modulus field $f$ must
be static and, together with the gauge field $A$, solve the following
set of the first order {\it self-dual} equations:
\begin{eqnarray}
\partial_if+\epsilon_{ij} f\,(\partial_j \theta + A_j) &=& 0
\label{SD1} \\
\kappa F_{12} + \frac{1}{2\kappa} f^2(f^2-v^2) - \rho_e &=&0
\label{SD2}
\end{eqnarray}
Here we have used Gauss' law to remove $\dot{\theta} +A_0$ in favor
of $\kappa F_{12}$.

Are there any vortex-like configurations solving
these self-dual equations?  First of all, the modulus field $f$ must vanish
at the center of a vortex in order for the Higgs field to be well-defined,
and this combined with  Eq. (\ref{SD1}) implies that a self-dual
vortex-like configuration always consists of anti-vortices only,
$\theta = -\sum_a {\rm Arg}\, (\vec{x}-\vec{q}_a)+\eta$ where $\vec{q}_a$
are the positions of the anti-vortices and $\eta$ is a single valued
function.

Then, we may combine the coupled first-order equations above to
produce a  single  second order equation for $f$ with sources at the
sites of anti-vortices.
\begin{equation}
\nabla^2 \ln f^2 -  \frac{1}{\kappa^2} [ f^2(f^2-v^2) - 2\kappa
\rho_e]  = 4\pi \sum_a  \delta(\vec{x}-\vec{q}_a) . \label{vortex}
\end{equation}
In the symmetric phase, for instance, we see that, far from the
solitons, the modulus field behaves exponentially
$f \propto e^{-\rho_e {\bf x}^2/4\kappa}$ so that there is no self-dual
configuration in the symmetric phase whenever $\rho_e <0 $.

(The solutions of this last equation in case the external charge density
vanishes were studied in detail before \cite{Hong}. In the symmetric phase,
there are q-balls and q-balls with vortices, while, in the asymmetric
phase, there are topological vortices.  One of interesting properties of
these solitons is that they carry the fractional spin and satisfy the
fractional spin statistics therein.)

\section{Vacuum Structure}

Next, let us consider the vacuum structure.
Of particular interest are those superselection sectors with homogeneous
vacuum states in it. One may naively expect to find a homogeneous ground
state in any given sector, but it is not difficult to see that there exist
only two cases where this is possible: either $\int f^2\,(\dot{\theta}+
A_0)=0$ or $\int F_{12}=0$.

The homogeneous ground state one finds in the first sector is just the
uniform magnetic field configuration, $F^e_{12}=\rho_e/\kappa$, without
any Higgs field expectation value, $\langle {\phi}\rangle =0$.  Since this
configuration saturates the self-dual equations trivially, this
symmetric homogeneous vacuum must be an absolute minimum within this
superselection sector.

{}From the form of the self-dual potential $U(f)$, however, one can see that
for sufficiently large and positive $\rho_e$, the potential is concave at
origin $|\phi| \equiv f/\sqrt 2 =0$, and the homogeneous symmetric
vacuum appears to be at a saddle point, unstable even perturbatively.
What are we missing? The uniform magnetic field $\rho_e/\kappa$ affects
the dynamics of the small perturbation profoundly, among other things.
To the first order in perturbation, the small deviation $\delta \phi=
\phi$ obeys the following equation with a nontrivial kinetic part..
\begin{equation}
-\partial_0^2\,\delta \phi=-(\partial_i -i A_{* i})^2\, \delta\phi
+U''(0)\,\delta \phi, \label{landau}
\end{equation}
where the gauge field configuration $A_{*i}$ satisfies $\partial_1 A_{*2}
-\partial_2 A_{*1}=\rho_e/\kappa$. To this leading order, the gauge
field fluctuation $A-A_*$ decouples from the Higgs mode, and actually vanishes
identically.

Note that a perturbative instability is possible if and only if the operator
on the right hand side of Eq. (\ref{landau}) possesses a negative eigenvalue.
But, this eigenvalue problem is just the well-known Landau level problem
with an extra ``potental'' term $U''(0)$. The eigenvalue spectrum of
the kinetic  part is $(2n+1)\,|\rho_e|/\kappa$ for all integers $n\ge 0$,
which implies that the operator is bounded below by $|\rho_e|/\kappa+U''(0)=
{|\rho_e|}/{\kappa}+[\,{v^4}/{4\kappa^2} -{\rho_e}/{\kappa}] \ge0 $.
Hence, the perturbative calculation  predicts that the homogeneous
symmetric vacua is indeed linearly stable. By the way, the infinite number
of zero modes that appear when $v^2=0$ (with positive $\rho_e$) is intimately
related to the vanishing self-dual energy-bound (\ref{SDE}).

\vskip 5mm
In the second sector, where Gauss's constraint is satisfied by
introducing an opposite scalar charge density, it is again possible to
obtain a homogeneous configuration that solves the field equations.
With $F_{12}=0$ everywhere, the constraint is then satisfied as
$a(f) \equiv \rho_e/f^2= \dot{\theta} +A_0$.  The effective energy density
of such configurations is then given by the sum of the scalar potential
and an  ``electrostatic'' contribution coming from the scalar kinetic term.
\begin{equation}
{\cal E}=\frac{\rho_e^2}{2 f^2}+U(f)=\frac{1}{8\kappa^2 f^2} (
f^2\,(f^2-v^2)-2\kappa\rho_e )^2.
\end{equation}
Searching for homogeneous vacua in this  second superselection sector
is now simply a matter of minimizing $\cal E$ with respect to $f$.

Once we find a local minimum at $f=u$, the next logical step is to
test its stability.  One convenient way is to check whether the
configuration saturates the self-dual bound. Since there is no net
magnetic flux, the self-dual bound of the energy functional vanishes,
and a minimum of $\cal E$ saturates the bound if and only if the value
of $\cal E$ there is {\it identically zero}. Later on, we will find
that this does not necessarily hold for all available minima.

On the other hand, we also have the option of carrying out a perturbation
to test the linear stability. For this, one again expands the field
equations
around the homogeneous vacuum. But unlike the previous case of symmetric
vacuum, the gauge field fluctuation does not decouple, and we find
a system of coupled linear partial differential equations with uniform
coefficients:
\begin{eqnarray}
[-\partial_0 \partial_0+\partial_i\partial_i+a^2(u)-U''(u)]\,\delta f+
2u\,a(u)\, \delta A_0 &=&0, \nonumber \\
\kappa \,\epsilon_{ij}\,\partial_i\,\delta A_j+
u^2\,\delta A_0+ 2u\,a(u)\,\delta f&=&0, \label{linear} \\
\kappa\, \epsilon_{ij}\,(\partial_0  \, \delta A_j-
\partial_j\, \delta A_0)+u^2\,\delta A_i&=& 0, \nonumber
\end{eqnarray}
where we used the gauge $\theta=0$.
A mode expansion with the frequency $w$ and the spatial
momentum $\vec{p}$ yields the following dispersion relation:
\begin{equation}
w^2=p^2+ \frac{1 }{2}\,\biggl({\cal E}''(u)+\frac{u^4}{\kappa^2}\biggr)
\pm\frac{1}{2}\sqrt{\biggl({\cal E}''(u)-\frac{u^4}{ \kappa^2}\biggr)^2
+16a^2(u)\,p^2}.
\end{equation}
Since ${\cal E}''$ is necessarily nonnegative at local minima, large
wavelength fluctuations are massive ones  with the masses squared given by
\begin{equation}
m_H^2(u)\equiv {\cal E}''(u),\qquad m_A^2(u) \equiv \frac{u^4}{\kappa^2}.
\end{equation}
These are associated with small fluctuations in the Higgs mode and in
the massive gauge field mode respectively. When $\rho_e=0$, the spin
of the Higgs particle of mass $m_H$ is zero, while that of the vector
particle of mass $m_A$ is one.  By the continuity, we expect this
aspect persists even with nonzero charge background. At the zero
momentum, Eq. (\ref{linear}) implies that the Higgs mode and the vector field
mode are decoupled, confirming that the spin of the Higgs particles
and the vector bosons are zero and one, respectively.  We also note that two
branches of the dispersion relation do not cross each other except at
$p^2=0$ when $m_A=m_H$.

A linear instability occurs if and only if $w^2$ becomes negative at some
nonnegative value of $p^2$. We
find the homogeneous vacuum of $f=u$ is linearly unstable if and only if
the following inequality is satisfied.
\begin{equation}
4a^2  > (m_H+m_A)^2. \label{unstable}
\end{equation}
This can be easily seen from the explicit forms of $w^2$ and $p^2$
when $w^2$ takes the minimum value, given the dispersion relation
above:
\begin{eqnarray}
p_*^2&=&-\frac{1}{16a^2}\,\biggl[\,{(m_H^2-m_A^2)^2}-
16a^4\biggr], \nonumber \\
w_*^2 &= &-\frac{ 1}{16a^2}\,\biggl[\,(m_H+m_A)^2-
4a^2\biggr]\,\biggl[\,(m_H-m_A)^2-4a^2\biggr] .
\end{eqnarray}
The $w^2_*$ can take a negative value if either $4a^2 >(m_H + m_A)^2$ or
$(m_H - m_A)^2 >4a^2$, but the latter implies $p_*^2   <0$,  and
thus is unphysical.

\vskip 5mm
In some cases, there exists a so-called magnetoroton mode that are
excitations with the least energy but at nonzero momentum. In the present
context, magnetorotons should appear whnever $p^2_* >0$. Depending on
whether $m_H> m_A$ or not, the roton mode occurs on the branch of the
Higgs mode or the vector boson mode. One thus naively expect that the
spin of the roton mode is zero or one, depending on whether $m_H>m_A$
or not. However, there is no rest frame for rotons and it is not clear
at this moment whether it is possible to separate the orbital and spin
angular momentum for rotons. As we shall see shortly, magnetorotons
in a stable vacuum exist for a limited range of $\kappa\rho_e$ and
that only for positive $v^2$.

Now let us list all possible minima of  $\cal E$
over various ranges of the parameters $v^2$ and $\kappa \rho_e$.
Extremizing $\cal E$, we find the following four solutions:
\begin{eqnarray}
u^2_\pm &\equiv& \frac{ v^2\pm\sqrt{v^4+8\kappa \rho_e}}{ 2}, \nonumber \\
\tilde{u}^2_\pm &\equiv&\frac{v^2\pm\sqrt{v^4-24\kappa \rho_e}}{6} .
\label{vacua}
\end{eqnarray}
Not all of these represent a physical vacuum. Depending on the parameters
of the theory, some of them become negative or even complex. Also,
a $u^2 \ge 0$ may correspond to a local {\it maximum} instead of a minimum.
Through a careful but elementary study of $\cal E$, one finds the
following homogeneous vacuum structure:

\vskip 8mm
\noindent
(1) $v^2 <0$, $0<\kappa \rho_e$;  $u_+$ is an absolute minimum and
saturates the self-dual bound. \hfill\break
(2) $v^2 <0$, $\kappa \rho_e <0$;  $\tilde{u}_+$ is an absolute
minimum  but does not saturate the self-dual bound. \hfill\break
(3) $v^2 >0$, $\kappa \rho_e >v^4/24$;  $u_+$ is an absolute minimum
and saturates the self-dual bound. \hfill\break
(4) $v^2 >0$, $v^4/24 >\kappa\rho_e >0$;  $u_+$ is an absolute
minimum and saturates the self-dual bound; $\tilde{u}_-$ is a local
minimum that does not saturate the self-dual bound. \hfill\break
(5) $v^2 >0$, $0> \kappa\rho_e > - v^4/8$;  $u_\pm$ are  two degenerate
absolute minima and saturates the self-dual bound. \hfill\break
(6) $v^2 >0$, $-v^4/8 >\kappa \rho_e$;  $\tilde{u}_+$ is an absolute
minimum but does not saturate the self-dual  bound.
\vskip 8mm

\noindent The cases where the equality rather than the inequality holds can be
understood as the degenerate limits of the cases in the above list.

Despite this long list of different cases, the linear
stabilities are more or less determined by the vacuum expectation
value $u$ alone. For example, the homogeneous vacua $u=u_\pm$,
whenever they are real, do saturate the self-dual bound ${\cal E}=0$,
implying that they are always stable under linear perturbations.
Indeed, an explicit calculation shows that
\begin{eqnarray}
 4a^2(u_+) &=& (m_H(u_+)-m_A(u_+))^2, \nonumber \\
4a^2(u_-)&=&(m_H(u_-)+m_A(u_+))^2.
\end{eqnarray}
Clearly both $u_\pm$ vacua are linearly stable, in view of the
instability criterion (\ref{unstable}). The above equation implies
$p_*^2(u_+) <0$ so that there is no roton in the $u_+$ vacuum.
Interestingly enough, $u_-$ vacuum is actually right at the borderline
of instability, and generically possesses a massless roton mode at $
p_*^2(u_-) = m_H(u_-)\, m_A(u_-) $.

The linear fluctuations around $\tilde{u}_\pm$ vacua are a bit more
complicated. First of all, neither $\tilde{u}_+$ nor $\tilde{u}_-$
saturates the self-dual bound since ${\cal E}(\tilde{u}_\pm)>0$
unless $\kappa\rho_e=0$ or $\kappa \rho_e =-v^4/8$: Even the linear
instability becomes a nontrivial issue. Using Eq. (\ref{vacua}),
we find the following behaviour of $w^2_*$ and  $p_*^2$
in terms of the vacuum expectation values $\tilde{u}_\pm$:
\begin{eqnarray}
p_*^2 &= & \frac{\tilde{u}^2_{\pm}\,(v^2-5\tilde{u}^2_{\pm})\,(v^2-2
\tilde{u}^2_{\pm})\,(v^2-\tilde{u}^2_{\pm})}{\kappa^2\,(v^2-
3\tilde{u}^2_{\pm})^2}, \nonumber \\
w_*^2 &= & \frac{4\tilde{u}^6_{\pm}\,  (2\tilde{u}^2_{\pm}-v^2) }{
\kappa^2\,(v^2-3\tilde{u}^2_{\pm})^2}, \label{tilde}
\end{eqnarray}
where we have traded off $\rho_e$ in favor of the vacuum expectation value
$\tilde{u}_\pm$.

In $\tilde{u}_+$ vacua for both cases (2) and (6), it is easy
to see that $w_*^2$ is always nonnegative, which again translates
into the linear stability. Are there roton modes? The answer is yes,
but only for case (6). $p_*^2$ is positive in this case, provided that
$\tilde{u}_+^2 \le v^2$. In terms of $\kappa\rho_e$, the corresponding
range is given by $-v^4/8> \kappa\rho_e \ge -v^4$.

Combined with the case (5) where a roton mode is found for $u_-$ vacuum,
this means that there exist a roton mode around a homogeneous asymmetric
vacuum only for theories with $ v^2>0 $ and $0 > \kappa\rho_e \ge -v^4$.
This roton is massless in $u_-$ vacuum while massive in $\tilde{u}_+$
vacuum. Note that the roton mode can occur on either branches of the
dispersion relation.

The only remaining vacuum to consider is that of the local minimum
$u^2=\tilde{u}_-^2$ that exists for $v^2>0$ and $v^4/24>\kappa \rho_e
>0$, namely for the case (4). Although it is a local minimum of the
effective energy density $\cal E$, it is always unstable.  Indeed, as
we vary $\kappa\rho_e$ within this range, $\tilde{u}_-^2$ interpolates
between $v^2/6$ and $0$, and Eq. (\ref{tilde}) then
implies that $w^2_*$ is always
negative while $p_*^2$ remains positive. Hence, this particular local
minimum is classically unstable against an inhomogeneous fluctuation (of
length scale $\sim \kappa/{v\tilde{u}_-}$), unlike any other case we
considered before. At this point it is not clear whether such
instability will lead to excitations on the homogeneous self-dual
vacuum $u^2=u_+^2$.

\section{Self-Dual Solitons}

Having studied the homogeneous vacuum structure in great detail, we
are now in position to ask what are possible self-dual soliton
solutions in the symmetric and the asymmetric phases.  Previous works
shows that in the models without the background charge density there
can be q-balls in the symmetric phase and vortices in the asymmetric
phase \cite{Hong}. Do they persist in our case? If so, how does the
background deform them? Are there other kind of solitons?
Specifically, we must investigate possible self-dual configurations
that asymptotically approach a given homogeneous vacuum.

Similarly to what we have seen in the study of homogeneous asymmetric
vacua, the answer depends on the parameters of the theory and also on
which homogeneous vacuum we consider. For instance, in $\tilde{u}_\pm$
vacua, the second self-dual equation (\ref{SD2})
implies a divergent total magnetic
flux, thus making a self-dual soliton impossible. Another case where
a self-dual soliton does not exist is the symmetric phase in theories
with negative $\rho_e$, as pointed out earlier.

Are there self-dual solitons in $u_\pm$ vacua? While a generic self-dual
configuration would not be rotationally symmetric, let us focus on the
rotationally symmetric solutions to gain an insight on the qualitative
features of the self-dual solutions. The ansatz for a rotationally
symmetric configuration around the origin is made of $f(r), \theta=
-n\varphi, A_i=A(r) \hat{\varphi}_i $. Rewriting the second order
form of the self-dual equation in terms of a new field $y\equiv \log
(f^2/|v^2|)$ and the dimensionless radial coordinate $s= r|v^2|/\kappa$,
we find a second order dissipative equation for
rotationally symmetric self-dual configurations:
\begin{equation}
\frac{d^2y}{ds^2}+\frac{1}{s}\frac{dy}{ds}+\frac{\partial V(y)}{\partial y}
= 2n \frac{\delta (s)}{s}, \label{master}
\end{equation}
where $y= \ln(f^2/|v^2|)$ and $V(y) =- e^{2y}/2 \pm e^y +2 \kappa
\rho_e y /v^4$, $\pm 1$ being the sign of $v^2$.

Let us first consider the $u_+$ vacuum. Note that $y_+= \ln (u_+^2/|v^2|)$
is always a local maximum of $V(y)$.  In Eq. (\ref{master}), a self-dual
anti-vortex in the $u_+$ vacuum starts at the asymptotic value $y_+=\log
(u^2_+/|v^2|)$ and smoothly evolves to $y=-\infty$, as $r$ decreases from
$\infty$ to $0$.  The dissipative term tends to increase the total
``energy'' as $r$ approaches the origin, and thus lets $f^2=|v^2| e^y$
reach zero at the center of the vortex even when $V$ has a linearly
increasing potential barrier toward $y=-\infty$ ($\rho_e <0$).

However, if the system starts out at the other possible asymptotic value
$y_-=\log (u^2_- /|v^2|)$ that exists in the case (5), the only nontrivial
solutions at large distances exhibit the following damped oscillatory
behaviour:
\begin{equation}
(y-y_-)\,  \propto \, \frac{1}{ \sqrt{s}}e^{\pm is \sqrt{V''(y_-)} } .
\end{equation}
Although this lets $f^2$ settle down to $u^2_-$ at large spatial distances,
the resulting configuration does not have a convergent total magnetic
flux,
\begin{equation}
\int^r_0 dr\, rF_{12} \sim \sqrt{r}e^{\pm ir|v^2|/\kappa\, \sqrt{V''(y_-)}
} .
\end{equation}
If we are considering a rotationally nonsymmetric soliton instead,
that will simply introduce an extra term $\partial^2_\phi f/s^2$ in the
equation above that cannot change this asymptotic behaviour. There is
simply no nontrivial self-dual configuration that asymptotes to the
homogeneous vacuum $u_-$. This is clearly related to the fact that
there are massless roton modes in this vacuum.

This leaves us with only two choices for the asymptotic state of self-dual
solitons: either the homogeneous asymmetric vacuum $f^2=u_+^2$  that exists
when $8\kappa \rho_e +v^4 >0$ or the the homogeneous symmetric vacuum
$f^2=0$  with $\rho_e \ge 0$. The second half of this section is mainly
devoted to understanding novel solitonic states that arise in these vacua
when the background $\rho_e$ is nontrivial. As we will see shortly, not
all of the self-dual solitons here survive the limit $\rho_e\rightarrow 0$.

First of all, in the asymmetric $u_+$ phase with $v^2>0$, the
self-dual solitons are rather similar to the zero background case.
Figure 1. shows the self-dual anti-vortex configuration for $n=1$,
$v^2>0$ and $\kappa \rho_e = v^4$.

\leavevmode
\epsfysize=3in \epsfbox{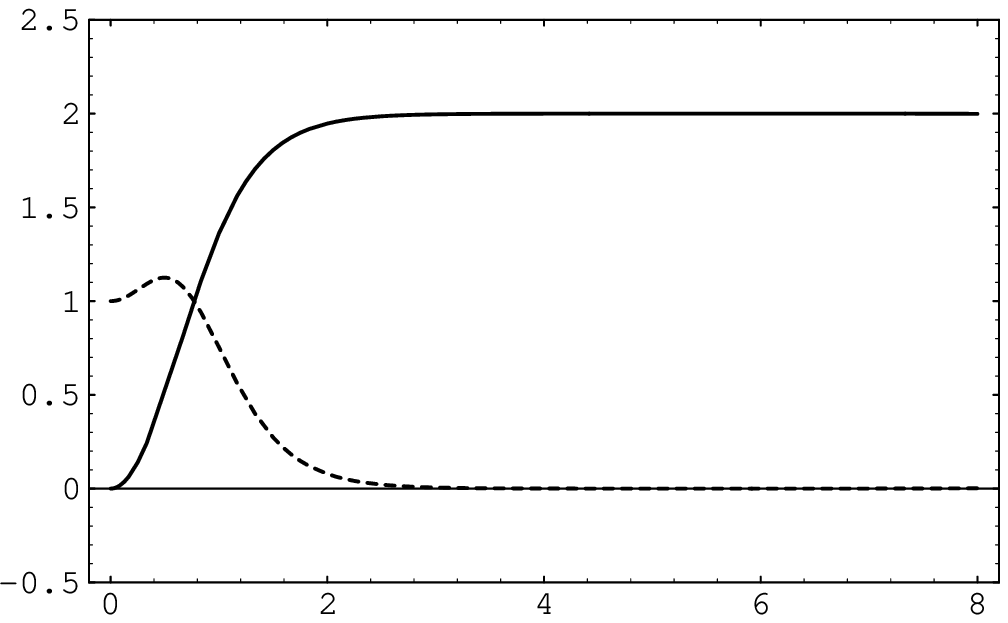}
\epsfysize=3in \epsfbox{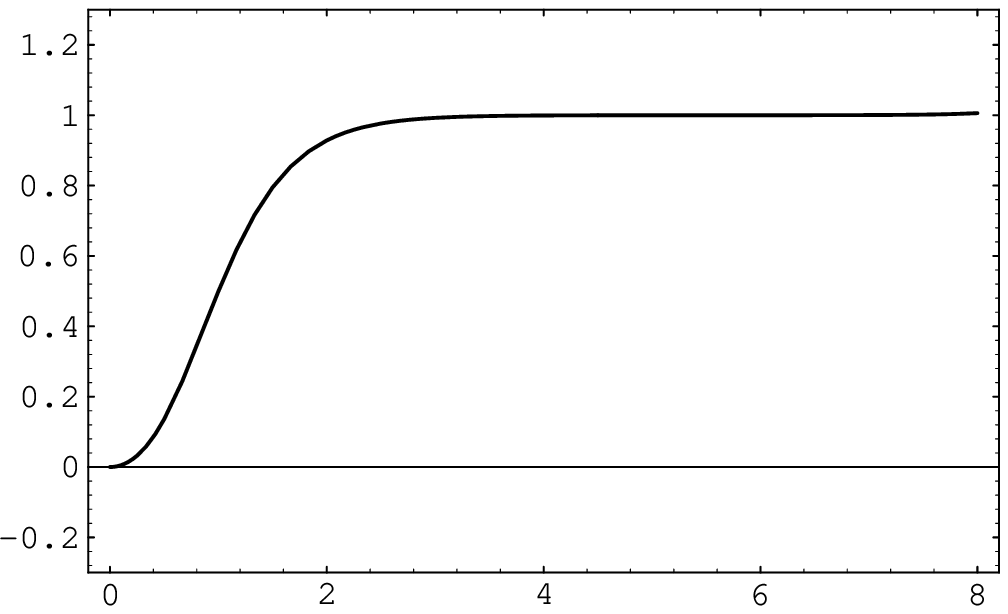}
\begin{quote}
\vskip -10mm
{\bf Figure 1:} {\small Plots of a typical rotationally symmetric
anti-vortex in the asymmetric phase that saturates the self-dual bound.
The first graph shows $f^2/|v^2|$ (solid line) and $\kappa F_{12}/\rho_e$
(broken line) as functions of the distance from the center. The second
graph is a plot of the flux (divided by $2\pi$) within finite distances
from the center. The radial distance is measured in terms of $\kappa/|v^2|$.
}
\end{quote}
\vskip 1cm

These self-dual anti-vortices do survive the limit $\rho_e \rightarrow 0$,
and become the ordinary topological solitons of the background-free theory.
However, if $v^2$ happens to be negative, an unexpected new type of
solitons arises, as we discuss below.

One interesting aspect of the self-dual bound here is that the energy
bound is proportional to the magnetic flux rather than to the absolute
value thereof. As a result, a unit self-dual anti-vortex in the $u_+$
vacuum of case (1) actually has a negative energy $-\pi |v^2|$.  Of
course, this does not mean that the homogeneous vacuum of case (1) is
unstable against the proliferation of anti-vortices, for whenever an
anti-vortex is created, a vortex must be also created to conserve the
total magnetic flux. In fact, as mentioned before Eq. (\ref{vortex}), and
due to the broken CTP, a vortex does not saturate its self-dual bound
$\pi |v^2|>0$, for it cannot solve the self-dual equations at the
vortex center.  Hence, a vortex-anti-vortex pair always costs more
energy than $\pi |v^2|- \pi |v^2|=0$, and the homogeneous asymmetric
vacuum is stable against this particular process. Even though
anti-vortices have negative rest energy, we see their kinetic mass
must be positive since the self-dual energy bound implies that slowly
moving anti-vortices should have a positive kinetic energy.

Still, one may wonder what happens to such negative energy solitons as
we take the limit $\rho_e \rightarrow 0$ where the self-dual bound of
the soliton energy is clearly positive. A useful equation to look at
in order to understand what happens, is the second self-dual equation
(\ref{SD2}) that can be solved for $f^2$ algebraically:
\begin{equation}
0\le \,2f^2=-|v^2|+\sqrt{v^4+8\kappa\rho_e-8\kappa^2 F_{12}}\,.
\end{equation}
The flux density $F_{12}$ must be bounded above by $\rho_e/\kappa$  to
maintain real $f$,  while the total
flux of the anti-vortex is quantized at a positive value $\int F_{12}
=2\pi$. Therefore, as $ \rho_e/\kappa \rightarrow 0$, the soliton core
must be spread out to infinity, leaving behind  a symmetric vacuum,
$f^2=0$.

What about the symmetric phase? Are there any negative energy q-balls,
possibly with an anti-vortex embedded inside. Again using the
the second self-dual equation, we can express the energy bound as
follows,
\begin{equation}
E-E_0 \ge \,\frac{v^2}{2}\,(\Psi-\Psi_0)=\frac{v^2}{2}\int d^2x\,\,
(F_{12} -\frac{\rho_e }{\kappa})=\frac{v^2}{4\kappa^2}\int  d^2 x
\, f^2\,(v^2-f^2) .
\end{equation}
For negative $v^2$, the right-hand-side is manifestly positive, while
the same could be true for positive $v^2$  provided that $f^2$ remains
smaller than $v^2$.

However, it is easy to see that, for a strictly positive $\rho_e$,
$f^2$ may grow larger than $v^2$. To see this, again consider
Eq. (\ref{master}), where the ``potential'' energy $V$ now has a unique
maximum at $y_+=\log (u_+^2/v^2) >0 $ but no minimum.
In this picture, a self-dual q-ball
would interpolate smoothly between $y=-\infty$ at $r=\infty$ and some
$y_0\neq -\infty$ at $r=0$, without the delta function source on the
right-hand side.  The only restriction on $y_0$ is that it be on the
left hand side of the maximum: $y_0 < y_+$, which translates into
$f^2(0) <u_+^2$.  Then, a q-ball may have a central region ($v^2 <f^2
<u_+^2$) of negative charge density ($F_{12}-\rho_e/\kappa <0$),
surrounded by an outer region ($0\le f^2 \le v^2$) that has a positive
charge cloud ($F_{12}-\rho_e/\kappa \ge 0$). In figure 2, two typical
shapes of q-balls are depicted, depending on whether $f^2(0)$ is
smaller or larger than $v^2$. In all figures in this section, the
parameters are chosen such that $\kappa\rho_e/v^4=1$.

\leavevmode
\epsfysize=3in \epsfbox{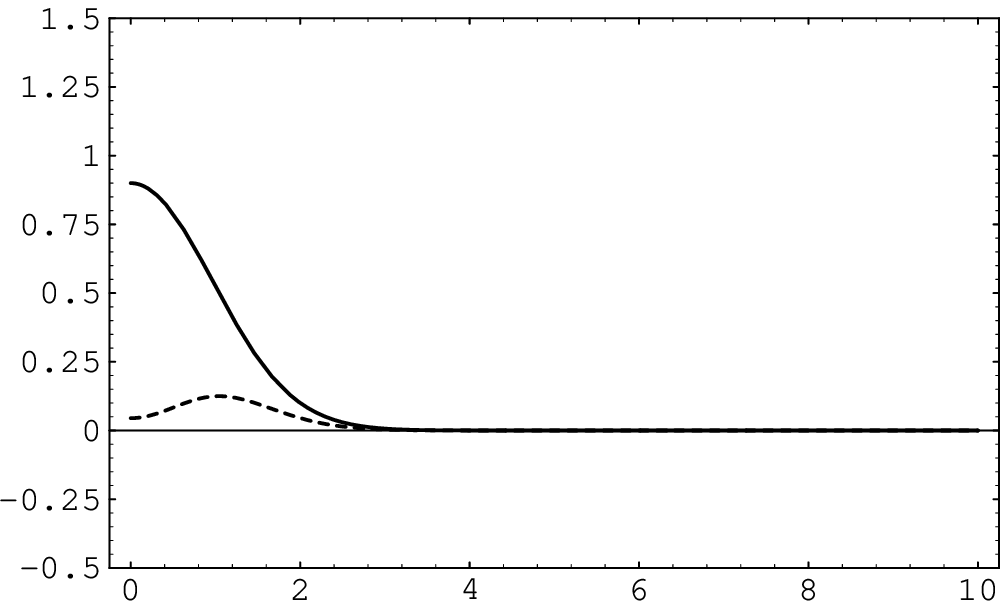}
\epsfysize=3in \epsfbox{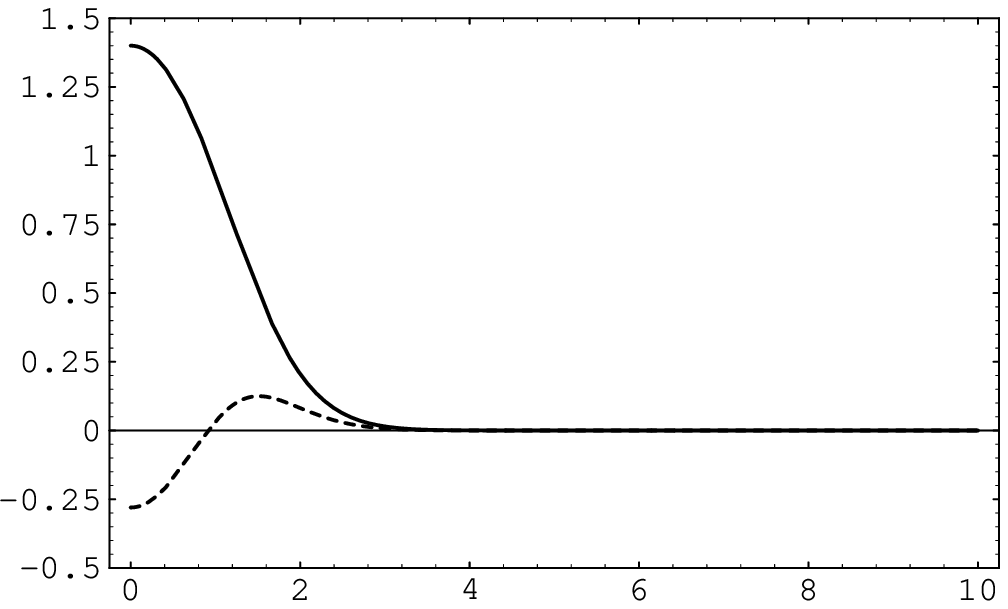}
\begin{quote}
\vskip -10mm
{\bf Figure 2:} {\small Plots of two q-balls in the symmetric phase
with $\kappa\rho_e/v^4=1$. Solid lines are $f^2/v^2$, and the
broken lines are the excess magnetic field, or equivalently, the
Higgs charge density  divided by $\kappa$, in the unit of $v^4/\kappa^2$.
The first graph corresponds  to $f^2(0)=0.9\,v^2$ while the second to
$f^2(0)=1.4\,v^2$.}
\end{quote}
\vskip 1cm

Furthermore, as $f^2(0)\rightarrow u_+^2$, the central region of
negative charge density becomes larger and larger, so that, in the strict
limit $f^2(0)= u_+^2$, it overtakes the entire plane such that
the corresponding self-dual configuration is simply that of the
homogeneous asymmetric vacuum $f^2(r)\equiv u_+^2$ with $F_{12}\equiv
0$. In other words, there exist self-dual q-balls of
arbitrary negative charge $\kappa\,(\Psi-\Psi_0)<0$, or equivalently of
arbitrary negative energy $E-E_0=(v^2/2)(\Psi-\Psi_0)<0$. This
is illustrated in figure 3.

\leavevmode
\epsfysize=3in \epsfbox{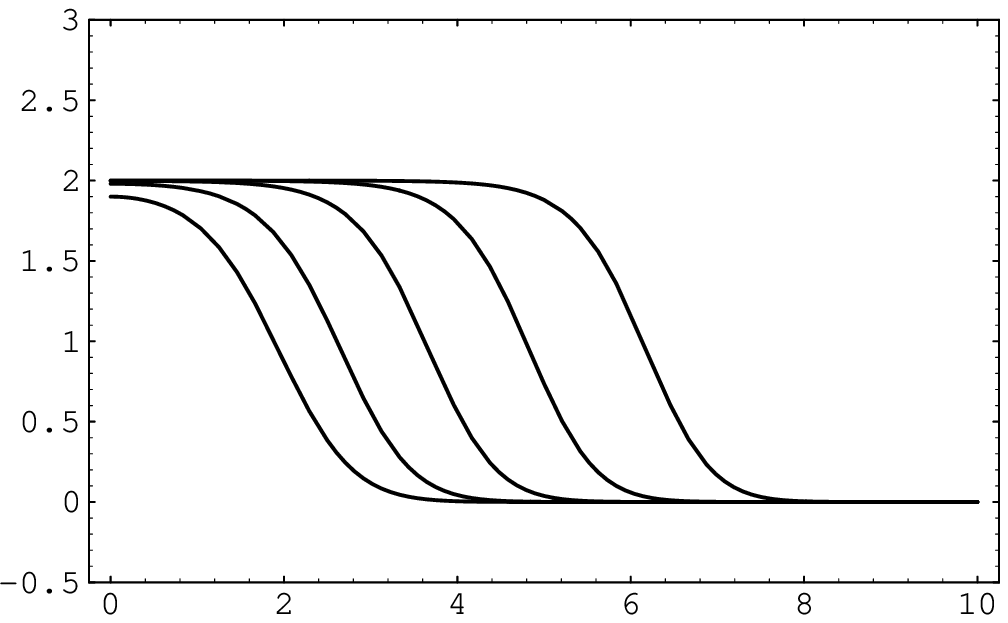}
\epsfysize=3in \epsfbox{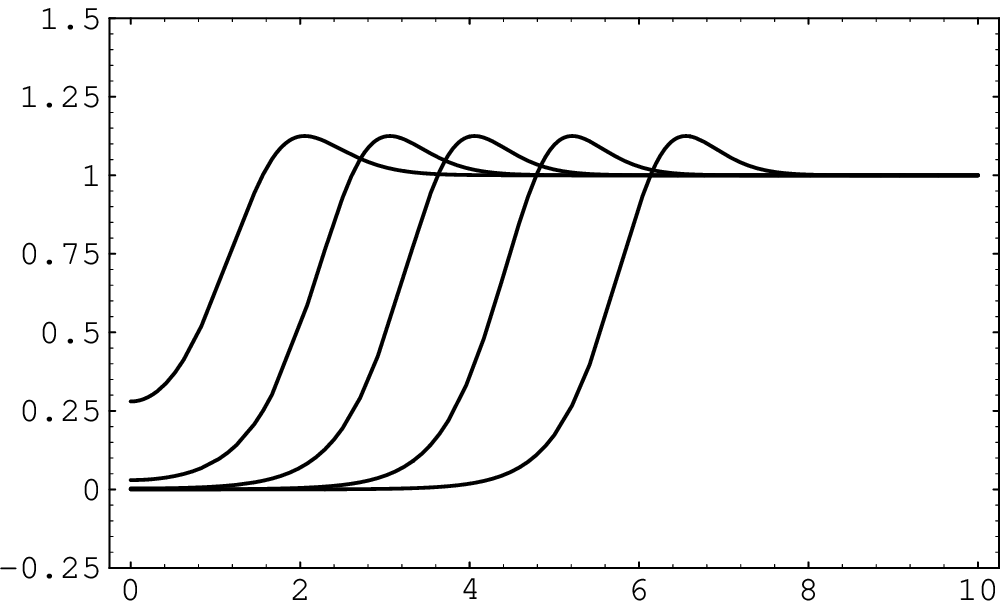}

\begin{quote}
\vskip -10mm
{\bf Figure 3:} {\small  Plots of $f^2/v^2$  and $\kappa F_{12}/\rho_e$
as the value of $f^2(0)$ approaches $u_+^2=2v^2$. As  $f^2(0)/u_+^2$
increases toward one from below, the central region of vanishing magnetic
field grows steadily (the second graph).}
\end{quote}

\vskip 1cm
In the other limit of $y_0\rightarrow -\infty$, the charge cloud of the
soliton remains manifestly positive everywhere, but the strength thereof
decreases indefinitely so that the net flux also decreases indefinitely.
The upshot is that we have a one-parameter family of rotationally
symmetric q-balls such that their total charge is {\it bounded from
above} by a finite positive quantity. Furthermore, for each net positive
charge $\kappa\,(\Psi-\Psi_0)>0$ allowed, there exist two different
solitons, depending on whether $f^2(0)$ is smaller than or larger than
$v^2$.

One interesting consequence of this is the degeneracy of the symmetric
vacuum. For a moderate value of $f^2(0)/v^2 >1$, there must exist
a nontrivial configuration of zero total charge, thus of zero net
magnetic flux and zero net energy. This is simply because there exists
a one-parameter family of solutions that interpolate between the
positive charge solution of $f^2(0)=v^2$ and  arbitrarily large
negative charge solutions that show up as $f^2(0)\rightarrow u_+^2$.
The zero-charge configuration looks like a small compact island of
broken phase enclosed by a cloud of positive charge that exactly
cancels the negative charge inside.
In figure  4, such a configuration is depicted when $\kappa  \rho_e
/v^4=1$.

\leavevmode
\epsfysize=3in \epsfbox{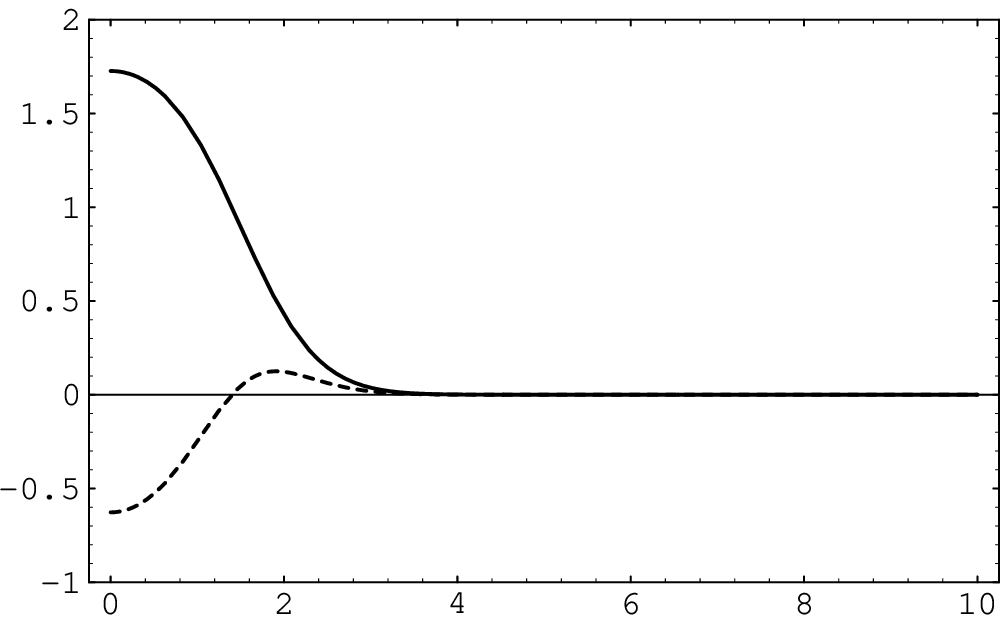}\hskip 1mm
\epsfysize=3in  \epsfbox{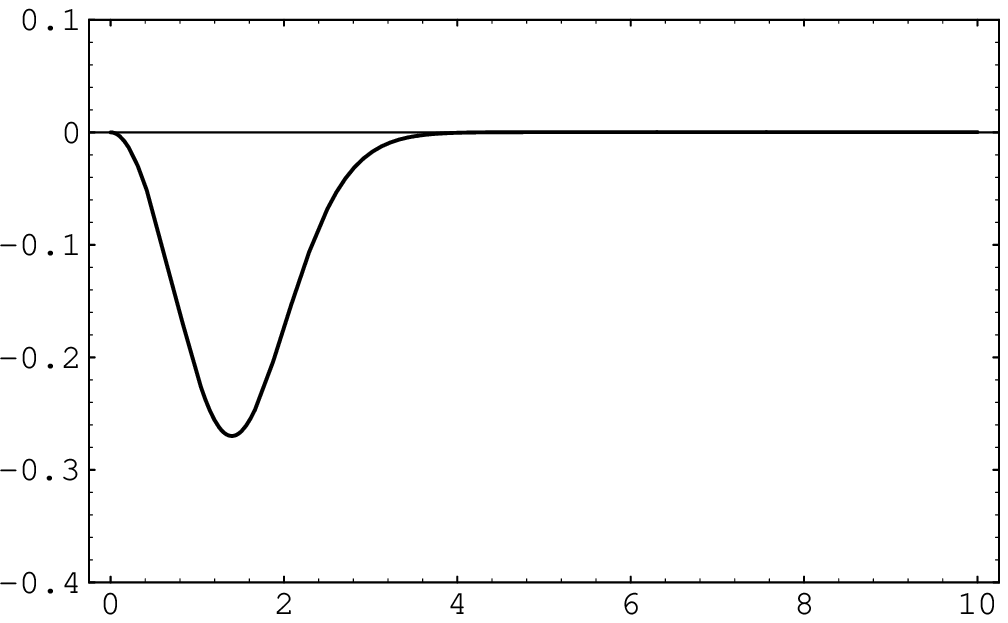}

\begin{quote}
\vskip -10mm
{\bf Figure 4:} {\small  Plots of a zero-energy solution. The second
plot shows the net flux (divided by $2\pi$)
within finite distances from the center of the
configuration as a function of $rv^2/\kappa$.}
\end{quote}

\vskip 1cm
It is not too difficult to convince ourselves that similar considerations
may be applied to q-balls with anti-vortices at the center. After fine-tuning
the behaviour of $f^2$ near the origin for each fixed anti-vorticity $n$, we
find that there exist rotationally symmetric solutions of arbitrary
negative net energy $E-E_0 <0$, and a unique solution of zero net energy
for each $n$. Such zero-energy configurations consist of a thick ring of
negative charge sandwiched between a positive charge core and a surrounding
outer region of positive charge cloud, such that the total charge is
exactly zero.

Hence, whenever both $v^2$ and $\rho_e$ are strictly positive,
the homogeneous symmetric vacuum is degenerate with infinite number of
rotationally symmetric self-dual configurations. Are there other, more
complicated, states of vanishing net charge that are also degenerate
with the homogeneous symmetric vacuum? One possible suggestion
is that these zero-energy ``solitons'' do not interact with each other so
that we may create many of them at once without costing any energy. This
possibility along with many interesting questions concerning general
multi-soliton configurations are postponed to a future study.

\section{Angular Momentum}

One of the more important aspects of Chern-Simon theories is the
fractional spin and statistics carried by the solitons. In view of the
fact that  the background charge density alters the
behaviour of the theory rather significantly, and in particular of
the explicitly broken CTP, it would be very interesting to find
out whether spin and statistics are modified. As  a first step
along this direction, let us consider the conserved angular (and
linear) momenta.

The naive angular momentum density obtained by the Noether theorem
contains a term, proportional to $\rho_e A_j$, that not only
cause divergences in all phases but is gauge-invariant only up to
a total derivative. Explicitly, the Noether angular momentum is
\begin{equation}
 J_{{\rm Noether}} = -\int d^2x \,\epsilon_{ij} x^i \biggl\{
\dot{f} \partial_j f + f^2(\dot{\theta} + A_0)(\partial_j \theta  +
A_j) -\rho_e A_j  \biggr\} .
\end{equation}
A conventional way of dealing with such problems is to isolate a
finite part by adding an appropriate manifestly conserved current, and
then justify that the final expression is a physically sensible one
\cite{Klee}.  However, we want to present an alternate and certainly
more elegant procedure by considering a similar system on a
two-sphere of radius $R$ in the limit $R \rightarrow \infty$.

A two-sphere has isometries in the form  of  $SO(3)$ rotations, and
there exist the corresponding conserved angular momentum 3-vector
$\bf L$.
Although the naive Noether expression for the density contains a
similarly problematic term as above, we may now perform a partial
integration to recover a manifestly gauge-invariant expression for
the density, since there is no longer a boundary to worry about:
\begin{equation}
 {\bf L} = -\int R^2d\Omega \, \biggl\{
\dot{f}\, ({\bf r} \times \nabla ) f
+f^2\, (\dot{\theta} +A_0)\, {\bf r} \times (\nabla \theta + {\bf A})
-\rho_e \,{\bf r}\, ({\bf r} \cdot \nabla\times {\bf A}) \biggr\}.
\label{3d}
\end{equation}
This  satisfies the usual $so(3)$ commutation relations. Here, $\bf r$ denotes
the coordinate vector of the three dimensional Euclidean space where the
spatial sphere is imbedded. In this notation, the actual spatial manifold
corresponds to $|{\bf r}|=R$. The 3-vector $\bf A$ must be tangent
to this two-sphere, since it represents the spatial gauge field on the
two-sphere. Finally, the total magnetic flux $\Psi = \int R^2d\Omega\, \,
{\hat{\bf r}} \cdot \nabla \times {\bf A}$ should be quantized in the unit
of $2\pi$ to ensure the gauge invariance.

Already, we can see the advantage of working on the sphere instead of
the plane: the last term is now proportional to $\nabla \times {\bf A}$,
thus manifestly gauge-invariant. Furthermore, given a {\it uniform}
radial magnetic field strength $B$, there is no $B R^4$ divergence in
the total angular momentum $\vec{L}$, due to cancellations between two
hemispheres, unlike $J_{\rm Noether}$ that has such a nasty
divergence in the symmetric phase.  In order to take the plane limit
$R \rightarrow \infty$, we imagine all interesting activities of the
system occurs within finite distances from the north pole.

First, let us consider the asymmetric phase where we have a uniform
Higgs charge density $-f^2(\dot{\theta}+A_0)=-\rho_e$ at large spatial
distances from the north pole. The vortices and the anti-vortices of
total flux $\Psi=2n\pi$, if distributed near the north pole, cause a
quadratic divergence $\Psi\,\rho_e R^2 $ in the third component $L^3$.
However, this should not be surprising at all. A similar divergence
arises if we consider an electrically charged particle on a large
two-sphere threaded by a uniform magnetic field $B$ of fixed strength.
The extra angular momentum $\sim eg\hat{\bf r}$, familar from the
studies of charge-monopole system \cite{coleman},
would diverges quadratically simply
because the total magnetic charge $g=B R^2$ diverges quadratically. In
the asymmetric phase, there exists the so-called Magnus force, between
the uniform Higgs charge density $-\rho_e$ and the localized fluxes,
which simulates the Lorentz force.

As long as we are concerned with the plane limit, we may as well remove
this divergence in each superselection sector of fixed $\Psi$. Then, we
find the following modified angular momentum, valid for the symmetric
phase  on the plane:
\begin{equation}
 J_{\rm asym} = - \int d^2x \biggl\{ \epsilon_{ij} x^i \biggl[
\dot{f} \partial_j f + f^2(\dot{\theta} + A_0) (\partial_j \theta +
A_j) \biggr] + \frac{\rho_e}{2} x^2 F_{12} \biggr\}.
\end{equation}
Note that the vestige of the Magnus force contribution still
remains in the last term. To see this more clearly, consider
a generic $\theta$ field configuration,
\begin{equation}
 \theta = - \sum_a (-1)_a {\rm Arg} (\vec{r}-\vec{q}_a(t)) + \eta,
\end{equation}
where $\eta$ is, as usual, single-valued. Then, the above angular momentum
can be rewritten, after an integration by part, as
\begin{equation}
J_{\rm asym} \ = -\int d^2x \,\,x^i\epsilon_{ij} \,\left\{
\dot{f} \partial_j f + \bigl[ f^2(\dot{\theta} + A_0) -\rho_e \bigr]
(\partial_j \theta  +  A_j)  \right\} - \pi\rho_e \sum_a
(-1)_a |\vec{q}_a|^2 . \label{asym}
\end{equation}
The last term that sums over the locations of vortices, is similar to
what one would get for a charged particle moving under a uniform
magnetic field. Vortices in our case feel the aforementioned Magnus
force instead, and this term represents the corresponding modification.
For a similar  expression in the pure Maxwell-Higgs systems, see
Ref.\cite{Klee}

In the large sphere limit, the linear momenta $P^i$ on the plane are
found from its  relation to the $SO(3)$ angular momentum via $L^1 = - R
P^2$ and $L^2 = R P^1$. Note that as long as the nontrivial configurations
are concentrated near the north pole there will be no quadratic
divergences in $L^1$ and $L^2$.  Alternately, we may extract $P^i$'s from
the plane angular momentum, for the latter transforms like $J \rightarrow
J + \epsilon_{ij} a^i P^j$ under an infinitesimal translation $x^i
\rightarrow x^i + a^i$. The two resulting expressions coincide with
each other. From the quadratically divergent part of the commutation
relation $[L^1, L^2]=iL^3$ in the plane limit, we find $[P^1,P^2] =
i\rho_e \Psi$. This modified form of the commutator is again due to
the Magnus force exerted by the Higgs charge density $-\rho_e$.
Here we have to use the commutation relation $[A_1 (x,t),A_2(y,t)] = i
\delta^2(x-y) / \kappa$.

In the symmetric phase, the field $f$ will be nonzero only near the north
pole so that at large distances from the north pole, there is a uniform
magnetic field $\hat{{\bf r}}\cdot \nabla \times {\bf A} =\rho_e/\kappa$.
Physically interesting configurations are then those with finite net
magnetic flux, $\Psi-\Psi_0 <\infty$ where $\Psi_0= 4\pi R^2 \rho_e/
\kappa$. Again nontrivial configurations are assumed to be within finite
distance from the north pole as $R \rightarrow \infty$.
In order to take the plane limit, it is sensible to subtract $0=R^3
\int d\Omega\, \rho_e{\bf r}$, and make the
angular momentum density to be concentrated near the north pole.
This effectively replaces $\hat{\bf r}\cdot \nabla \times {\bf A}$ by
$\hat{\bf r} \cdot \nabla \times {\bf A} - \rho_e/\kappa$ in the last
term in Eq. (\ref{3d}).

Now, $L^3$ diverges quadratically in the plane limit as $R^2 \rho_e \,
(\Psi - \Psi_0)$. In the symmetric phase, the origin of this divergence
is even more transparent, for the system is basically that of charged
particles (q-balls) of total charge $\kappa\, (\Psi-\Psi_0)$, moving in
a large two-sphere that surrounds total magnetic charge $R^2\rho_e /\kappa$
at the center. In each superselection sector, we may again remove the
quadratic divergences to obtain the modified angular momentum, valid for the
symmetric phase on the plane.
\begin{equation}
 J_{{\rm sym}} = - \int d^2x \,\biggl\{
\epsilon_{ij} x^i \biggl[ \dot{f}\partial_j f  +
f^2(\dot{\theta} + A_0) (\partial_j \theta + A_j) \biggr]
+ \frac{ \rho_e}{ 2\kappa} x^2 \,(\kappa F_{12} -\rho_e) \biggr\} .
\end{equation}
\noindent
We can find the linear momenta similarly as in the asymmetric phase, and
can show that they satisfy a nontrivial commutator $[P^1,P^2]= i\rho_e
(\Psi - \Psi_0)$. This last expression may be rewritten, using
Gauss's constraint, in terms of the Higgs charge density and the background
magnetic field $\rho_e/\kappa$ of the symmetric phase: $[P^1,P^2]
=-i\,(\rho_e/\kappa) \int d^2x\,f^2\,(\dot{\theta}+A_0)$. Written this way,
the nontrivial commutator can be naturally attributed to the
velocity-dependent Lorentz force on the q-balls, exerted by the homogeneous
magnetic field $\rho_e/\kappa$.

\vskip 5mm
Now that we derived the correct angular momenta in the presence of the
uniform background fields, it is time to consider how these abstract
expressions translates into angular momenta of specific solitons.
For simplicity, we shall consider single static solitons.

Let us start with topological anti-vortex in the asymmetric phase.
Since we can safely exclude the center of the soliton at $x^i=q^i$
from the space integration of Eq. (\ref{asym}) without affecting the
value of the angular momentum, we may introduce a vector potential
$\bar{ A}_j = \partial_j \theta +  A_j$, and rewrite the $J_{\rm
asym}$ as an integral over
$R^2_* \equiv R^2-\{\vec{q}\}$:
\begin{equation}
 J_{{\rm asym}} = -   \int_{R^2_*} d^2 x \;  \epsilon_{ij}
x^i \,\biggl\{\dot{f}\partial_j f - \kappa\bar{A}_j \epsilon_{kl}
\partial_k \bar{A}_l \biggr\}  - n\pi \rho_e |\vec{q}|^2 .
\end{equation}
For the static self-dual solutions (and also for rotationally symmetric
solutions), $\partial_i \bar{A}_i = 0$ and  he integrand becomes a total
divergence:
\begin{equation}
J_{{\rm asym}} =  \kappa \int_{R_*^2} d^2 x\,\,\partial_j\,\left\{
{1\over 2}x^j (\bar{A}^k\bar{A}_k) - \bar{A}_j (x^k \bar{A}_k)\right\}
 - n\pi \rho_e |\vec{q}|^2 . \label{asym2}
\end{equation}
Note that $\bar{A}$, as a gauge invariant quantity, must be exponentially
small at large distances so that the only boundary to speak of is at
the center. If the anti-vortex has $n$ topological charge  (i.e.,
if the total magnetic flux is $2\pi n>0$), $\bar{A}_j$ near the center is
dominated by $\partial_j\theta \simeq n\epsilon_{jk}x^k/r^2$, and the
angular momentum is easily  obtained as follows,
\begin{equation}
J_{{\rm asym}} = -\pi \kappa n^2+\pi n\,(-\rho_e)\, |\vec{q}|^2 .
\end{equation}
In the limit $\rho_e \rightarrow 0$, this reproduces the expected results
found in Ref.\cite{Hong,Min}. The last term obviously reflects the Magnus
force on the anti-vortex exerted by the background Higgs charge density
$-f^2\,(\dot{\theta}+A_0)=-\rho_e$. Clearly, the above expression is
true also for rotationally symmetric vortices. Thus, we have shown vortices
and anti-vortices in Chern-Simons systems with nonzero magnetic flux carry
nonzero spins $-\pi\kappa n^2$ that are independent of $\rho_e$.

In the self-dual case, we can have  static multi-anti-vortex configurations
where anti-vortices do not overlap each other. Eq. (\ref{asym2}) can then
be explored much further to gain an understanding of slowly moving
vortices and thus of their statistics. (See Ref.\cite{Min} for the case
$\rho_e=0$.) We hope to return to this topic in near future.

In the symmetric phase, on the other hand, we introduce  $\tilde{A}_i
\equiv \partial_i \theta + A_i + \epsilon_{ij} x^j \rho_e/ (2\kappa)$.
Using Gauss's constraint, we again find that the angular momentum
for a static self-dual configurations can be written as an integral
of a total divergence,
\begin{equation}
J_{{\rm sym}} =  \kappa \int_{R^2_*} d^2r \epsilon_{ij}
x^i \tilde{A}_j \epsilon_{kl} \partial_k \tilde{A}_l
=\kappa \int_{R^2_*} d^2  x\,\,\partial_j\,\left\{
\frac{1}{ 2}x^j (\tilde{A}^k\tilde{A}_k)
 - \tilde{A}_j (x^k \tilde{A}_k)\right\} .
\end{equation}
As we have seen earlier in a more general context, the angular momentum
in the symmetric phase must also contain a term $\sim |\vec{ q}|^2$ that
originates from the Lorentz force exerted by the uniform
magnetic field $\rho_e/\kappa$. However, this expression, compared to
the analogue in the asymmetric phase, does not seem to contain such
a term. The crucial observation that resolves this apparent puzzle is
that, unlike $\bar{A}$ above in the asymmetric phase, $\tilde{A}$ exhibits
a powerlike behaviour at large distances so that we need to worry
about the contribution from the boundary at infinity. In fact,
if the soliton does not contain any vortex or anti-vortex at the
center and thus is a pure q-ball, the only boundary is at infinity.
\begin{equation}
J_{{\rm sym}} = \kappa \oint_{\infty}  dl  \biggl\{
\frac{r}{2}\, (\tilde{A}^k \tilde{A}_k)
-  \frac{(x^k \tilde{A}_k)^2}{r}\biggr\}.
\end{equation}
For a single, rotationally symmetric q-ball of net magnetic flux $2\pi
\alpha$, in particular, the self-dual equations (\ref{SD1}) and
(\ref{vortex}) imply the following asymptotic behaviour,
\begin{equation}
\tilde{A}_i \rightarrow \epsilon_{ij} \biggl\{ -\alpha \frac{x^j-q^j
}{|\vec{x}-\vec{q}|^2} + \frac{\rho_e}{ 2\kappa} q^j \biggr\},
\end{equation}
provided that the soliton is centered at $x^i=q^i$.
After a careful usage of this limiting form, we find the total angular
momentum  of a self-dual q-ball of net magnetic flux $2\pi \alpha$
at $x^i=q^i$:
\begin{equation}
J_{{\rm sym}} = \pi \kappa |\alpha|^2 -
\pi  \kappa\alpha\, \frac{\rho_e}{\kappa} \,|\vec{q}|^2 .
\end{equation}
The last term is exactly the position dependent term we need,
which again shows that our angular momentum is sensible.

In this section, we derived appropriate expressions of angular
momentum in each of the symmetric and asymmetric phases,  and
evaluated them on self-dual q-balls and anti-vortices. While
the angular momentum acquires a new position dependent terms that
reflects the interaction between the solitons and uniform
fields of the ground states, it also include the usual fractional
spin that is independent of the background. Those additional position
dependent terms, which are proportional to $\rho_e$, arise because the
slowly moving solitons feel a velocity-dependent force exerted by the
uniform fields of the surrounding vacuum.

\section{Conclusion}

We studied the characteristics of a self-dual Chern-Simons Higgs model
coupled to a background electric charge density. We found a rich vacuum
structure, and subsequently tested the classical stability of homogeneous
vacua along magnetoroton modes. In certain stable vacua, novel self-dual
solitonic configurations are found and studied in detail. Finally, the
divergent Noether angular momentum is successfully modified to a sensible
and finite expression. From this formula, we find that the soliton
angular momentum is corrected by an effective Lorentz force, and acquires
a term quadratic in the position as well as the usual
(background-independent) fractional spins.

As the structure of the theory is very rich, we were not able to
cover even the classical aspects of the theory completely. For instance,
there are cases where no self-dual solution exists: Although we
still expect to find non-self-dual solitons in such cases, practically
nothing is known about them other than the spin. Another important
topic which is not covered here is the matter of the spin-statistics.
Since our solitons carry nonzero fractional spin, we expect there is
nontrivial statistical interaction between solitons. It would be
interesting to find out whether the statistical interaction follows
the naive expectation that it is made of the spin contribution and the
background Magnus/Lorentz force contribution. The  study of the classical
dynamics of slowly moving anti-vortices in the asymmetric vacuum
may lead to a better understanding in this regard. Finally, it
is not clear to us at all how our self-dual model could be extended
to have $N=2$ or $N=3$ supersymmetry, which is expected from self-dual
models in general \cite{Susy}.

Quantum aspect of our theory should be also quite rich. Especially, the
understanding of the infinite vacuum degeneracy of the symmetric
phase poses a challenge.

\vskip 1cm
\centerline{\bf Acknowledgement}
\vskip 5mm
\noindent
We wish to thank Prof. Choonkyu Lee for drawing our attention to
this model. This work is supported in part by the US Department of
Energy. K.L. is also supported in part by the NSF Presidential Young
Investigator program and the Alfred P. Sloan Foundation.

\end{document}